# MR PRISM – Spectral Analysis Tool for the CRISM

Adrian J. Brown*[a], and Michael Storrie-Lombardi[b]

[a] SETI Institute, 515 N. Whisman Rd, Mountain View, CA 94043
[b] Kinohi Institute, Pasadena, CA 91101

## ABSTRACT

We describe a computer application designed to analyze hyperspectral data collected by the Compact Infrared Spectrometer for Mars (CRISM). The application links the spectral, imaging and mapping perspectives on the eventual CRISM dataset by presenting the user with three different ways to analyze the data.

One of the goals when developing this instrument is to build in the latest algorithms for detection of spectrally compelling targets on the surface of the Red Planet, so they may be available to the Planetary Science community without cost and with a minimal learning barrier to cross. This will allow the Astrobiology community to look for targets of interest such as hydrothermal minerals, sulfate minerals and hydrous minerals and be able to map the extent of these minerals using the most up-to-date and effective algorithms.

The application is programmed in Java and will be made available for Windows, Mac and Linux platforms. Users will be able to embed Groovy scripts into the program in order to extend its functionality. The first collection of CRISM data will occur in September of 2006 and this data will be made publicly available six months later via the Planetary Datasystem (PDS). Potential users in the community should therefore look forward to a release date mid-2007.

Although exploration of the CRISM data set is the motivating force for developing these software tools, the ease of writing additional Groovy scripts to access other data sets makes the tools useful for mineral exploration, crop management, and characterization of extreme environments here on Earth or other terrestrial planets. The system can be easily implemented for use by high school, college, and graduate level students.

# 1. INTRODUCTION

## 1.1 The CRISM Instrument on MRO

CRISM is a VNIR imaging spectrometer sensitive to light from 0.362 to 3.92 μm. In hyperspectral mode, it covers this region with a spectral spacing of 6.55 ηm. At MRO's nearly circular primary mapping orbit (255km x 320km) in hyperspectral mode CRISM will have a pixel size on the ground of 15-19m. CRISM is also intended to construct a global map using 70 chosen channels with a spatial pixel size of 100-200m on the ground (Murchie *et al.*, 2004).

CRISM has the ability to view the surface of Mars at multiple angles during data collection, enabling it to separate effects due to atmospheric absorption bands and those due to surface absorption features. Particular attention has been paid to calibration of the instrument prior to launch, meaning it has the potential to deliver outstanding signal to noise characteristics. These capabilities make it the most powerful imaging spectrometer sent into orbit, and place it far in advance of other spectrometers sent to Mars (Table 1).

CRISM covers the extremely important short wave infrared (SWIR: 2-2.5 μm) part of the spectrum, which corresponds to energies of combination and overtone bands of the hydroxyl anion ($OH^-$). The central wavelength of these absorption bands varies according to the cation bonding with the hydroxyl ion, for example $Al^{3+}$(2.2μm), $Fe^{2+}$ and $Fe^{3+}$ (2.25μm) and $Mg^{3+}$ (2.3μm) (Hunt 1979, Clark 1990). In hyperspectral mode CRISM will have 40 bands covering the SWIR region. This gives CRISM the ability to not only determine whether hydroxyl is present but also the nature of the cation bonding to the hydroxyl. Mineral signatures in this area of the spectrum are often diagnostic for particular hydroxyl minerals, including minerals such as amphiboles, phyllosilicates and hydroxylated sulfates like jarosite and alunite (Clark 1990, Brown 2006a). Carbonates also display a diagnostic absorption band at around 2.3 μm, but it is often masked by other absorption features (Brown *et al.*, 2004).

This project is designed to study the absorption bands of various sulfate minerals in order to learn more about their origin, cation and hydration states by studying several low hydration Ca and Mg sulfates. This information will be used to interpret future Mars data and expand the understanding of the Martian hydrological cycle.

| *Mission* | *Instrument Name* | *Operational Dates* | *Spectral coverage* | *Spatial Resolution* | *No. Spectral Bands* | *Reference* |
|---|---|---|---|---|---|---|
| Mariner 4 | IRS | 14 Jul 1965 | VIS | - | - | - |
| Mariner 6/7 | IRS | 1969 | 1.9-14.4μm | - | 4 channels | (Kirkland *et al.*, 2000) |
| Mariner 9 | IRIS | 1971 | 5-50μm | - | - | (Hanel *et al.*, 1972) |
| Viking | IRTM | 1977-1982 | 5-50μm | 20km | - | (Hunt *et al.*, 1980) |
| Phobos | ISM | 1989 | 0.77-3.14μm | 20km | - | (Bibring 1989) |
| Mars Global Surveyor | TES | 1997-present | 5-50μm | 2km | 7 channels | (Christensen 2000) |
| Mars Odyssey | THEMIS | 2001-present | VIS and TIR Multispectral | 20m | 10 VIS, 5 TIR channels | (Christensen 2004) |
| Mars Express | OMEGA | 2004-present | 0.35-5.1μm | 3-5km | 484 channels | (Bibring *et al.*, 2005) |
| Mars Recon Orbiter | CRISM | 2006-present | 0.36-3.92μm | 15m-200m | 570 channels | (Murchie *et al.*, 2004) |

Table 1 – Spectrometers designed to study surface minerals that have returned data from Mars orbit.

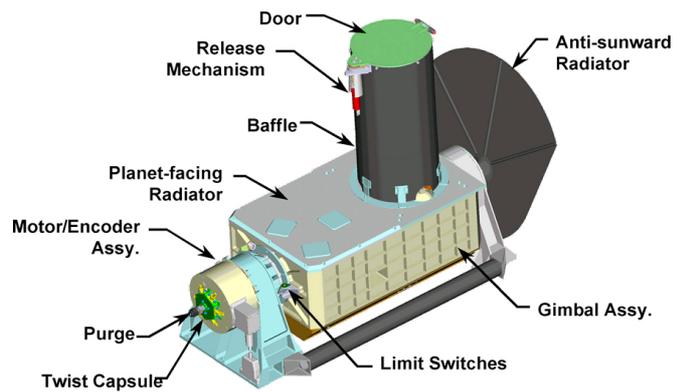 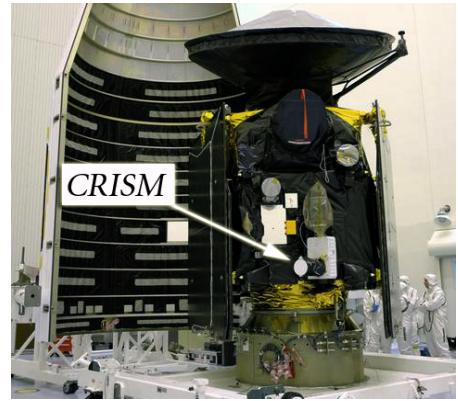

Figure 1 *(left)* – The Compact Reconnaissance Infrared Spectrometer for Mars - CRISM  *(right)* Mars Reconnaissance Orbiter being encapsulated at Kennedy Space Center.

## 2. MR PRISM – A NEW DATA ANALYSIS TOOL

**2.1 Motivation and Development**

As part of postdoctoral research being carried out with Dr. Janice Bishop (a Co-I on the CRISM team), the lead author has had the chance to develop a computer application suited for obtaining CRISM images from their repository at APL and analysing them on a local computer. The application is called Mars Reconnaissance PRISM – or MR PRISM for short.

The program has been developed from the ground up as a standalone cross platform Java application. It has been tested on Windows, Mac and Solaris platforms so far. The program is still in testing mode with researchers at the SETI Institute. It is designed as a no-cost alternative to commercial image analysis programs such as Geomatica, ArcMap and ENVI/IDL, which are not suited well to the CRISM hyperspectral dataset (Table 2).

MR PRISM operates in three modes – spectral, image and map analysis mode. Typically during an analysis session, a user will identify CRISM swathes on the map mode, request the data be loaded across the internet and placed in to the image mode, and as required will transfer spectra across to the spectral analysis mode. Because it is designed to handle CRISM data, MR PRISM aims to be a 'one-stop shop' for the CRISM analyst. These modes are shown in Figure 2.

MR PRISM has extensive support for automated routines, including the ability to run long lived processes in the background while the user completes other tasks. This will be important for running the automated absorption band analysis routines discussed earlier.

MR PRISM is designed to be extendable by end users. Embedded within the application is a scripting language similar to Java or Javascript called "Groovy". This allows a user to run commands within the application *while the application is running*, similar to 'EASI' on *Geomatica* or 'IDL' on *ENVI*. MR PRISM loads Groovy scripts on startup that are loaded into menus automatically. These sample scripts are shipped with MR PRISM.

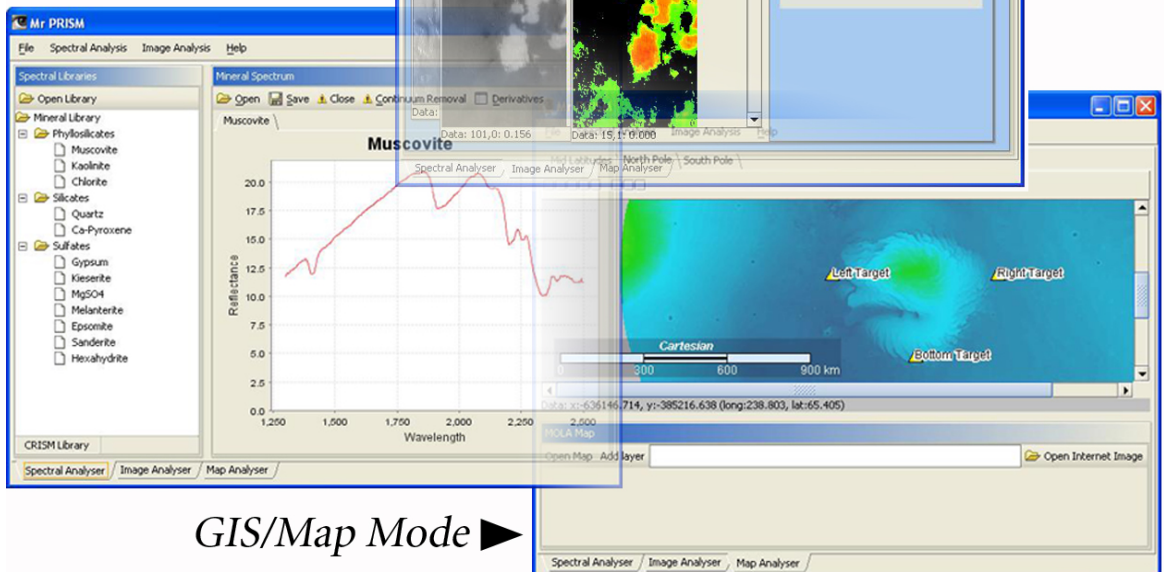

Figure 2 – MR PRISM in operation – (uppermost) Image Analysis (lower left) Spectral Analysis (lower left) Map (GIS) Analysis

Work is currently ongoing to document all the features of MR PRISM on a Wiki internet website similar to *Wikipedia*. It is anticipated that an initial version of MR PRISM will be available for external beta testing along with the first public release of CRISM data, in mid-2007.

In the early phases of CRISM data analysis, it is intended that MR PRISM will be used to produce maps of hydroxyl bearing minerals in the polar regions (Brown *et al.*, 2006). This analysis is easier in MR PRISM because the program has the ability to show CRISM coverage in polar stereographic projection (see Figure 2).

Future development of the MR PRISM program will concentrate on features needed by CRISM analysts, be they hyperspectral savvy spectroscopists or just have a passing interest in spectroscopy. Some lines of development include:

- integration of other MRO datasets, such as HiRISE, CTX and SHARAD data
- integration of CRISM atmospheric removal processes
- integration of CRISM backplane data, including MOLA and thermal emission maps

| Software | Hyperspectral analysis of IMG files | Built-in scripting language | Open source code | Cross platform support | Reads CRISM backplanes | Inbuilt GIS tool | Website |
|----------|---|---|---|---|---|---|---|
| MR PRISM | X | X | X | X | P | X | abrown.seti.org |
| ENVI/IDL | X | X |   | X |   |   | rsinc.com |
| Geomatica | X | X |   | X |   |   | pcigeomatics.com |
| JMars |   |   |   | X |   | X | jmars.asu.edu |

Table 2 – Comparison of key analysis features of MR PRISM to ENVI and Geomatica. An 'X' indicates an existing features and 'P' is a planned feature.

Following the release of CRISM data and MR PRISM, funding of this investigation will allow on going support for community enquiries and requests for MR PRISM features expected following its public release.

## 3. METHOD AND RESULTS

### 3.1 Demonstration

In order to demonstrate the capabilities of MR PRISM, we obtained data from the *Observatoire pour la Minéralogie, l'Eau, la Glace et l'Activité* (OMEGA) spectrometer on Mars Express (MEX). OMEGA has been acquiring visible-near infrared (VNIR: 0.4-5.0μm) reflectance spectra of Mars since late 2003 (Bibring *et al.*, 2006). We obtained a swathe of OMEGA data from scene 3 of orbit 1012, from Dr. F Poulet, a member of the OMEGA Science team. This scene covered a part of the Vastitas Borealis Formation (VBF) and the north polar cap ice deposit. The data had been atmospherically corrected using a simple scheme where the atmospheric signature is approximated using measurements taken from the top and base of Olympus Mons (Bibring *et al.*, 2005, Poulet *et al.*, 2006).

We have previously analyzed OMEGA data in the Athabasca Valles region using a computer program written in IDL, and using the ENVI software (Brown 2006b-e). Here we follow a simpler approach to show how a classification system works on MR PRISM.

In order to separate water ice deposits from soil, we wrote a short Groovy script that conducted a simple band ratio, using OMEGA band 58 (1.76 microns) divided by band 61 (1.80 microns). The resulting image is shown side by side with a grayscale image using band 342 (1.00 micron). Note that OMEGA bands are not sequential – the Visible bands are placed after the NIR bands.

The resulting band ratio image has been treated with a simple look-up table in the image shown on the left side of Figure 3. This makes the polar water ice deposits appear red and surrounding soil appear green. It also has the unfortunate effect of enhancing a calibration issue with this image as a vertical noisy streak through the image – this effect is seen in all bands of the NIR spectrometer in this image and is common in OMEGA images.

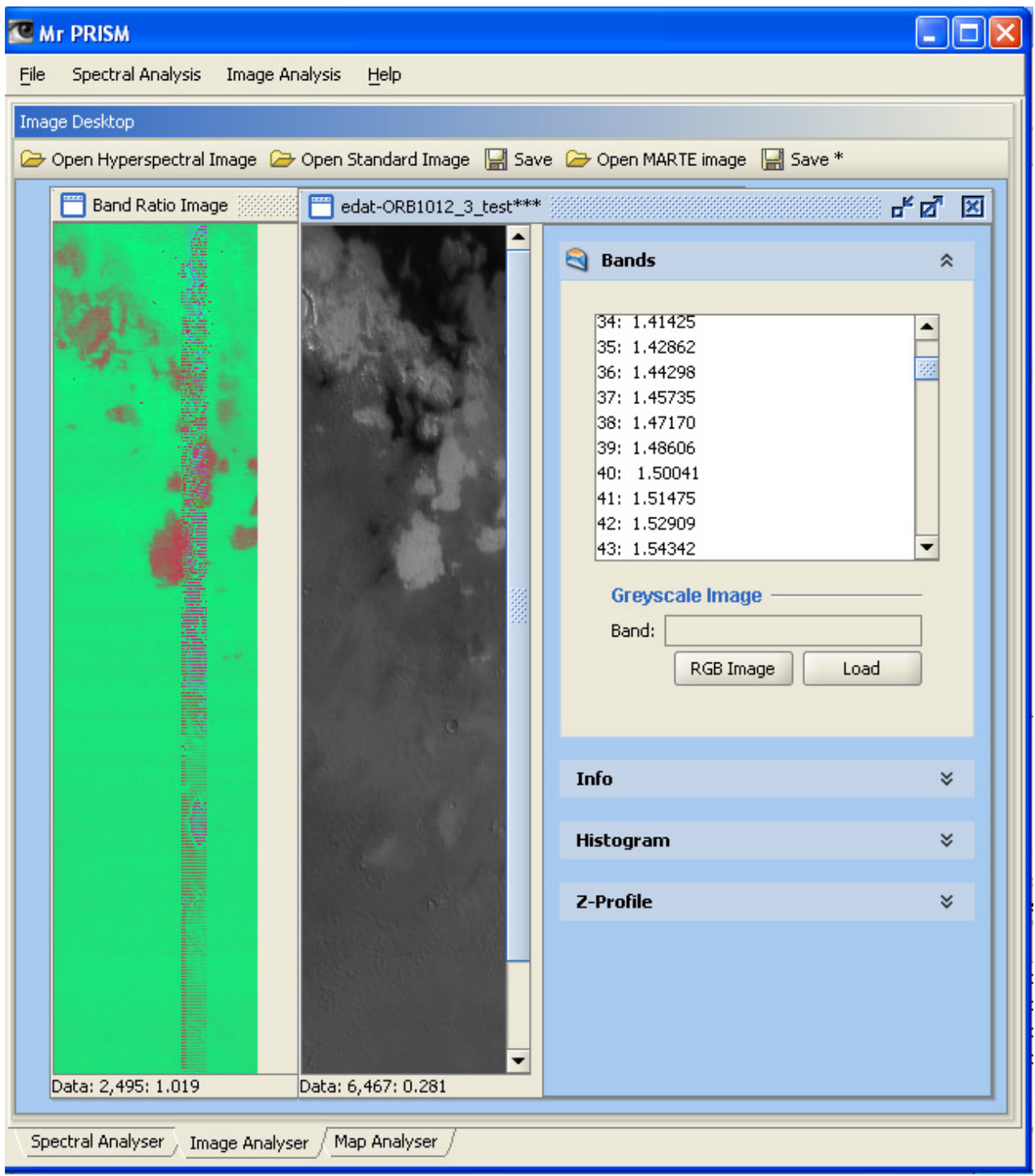

Figure 3 – OMEGA data from scene 3 of orbit 1012 – (left) a band ratio separating water ice deposits (in red) from soil (in green) (right) a grayscale image in the visible using band 342.

## 4. DISCUSSION

The example demonstrated here is somewhat trivial – it demonstrates simply that MR PRISM can be used to conduct simple band ratios that are commonly used in the analysis of multiband and hyperspectral analysis. It is proposed by the CRISM team that band ratio images will be made available to the interested public in order to assist the planetary community in making low level assessments of sites on Mars for such things as hydrated minerals, sulfates, and water clouds. Now the challenge is to take the next step in complexity.

Future developments for MR PRISM include a Bayesian analysis engine, the capacity to handle atmospheric correction routines provided by the CRISM team, and eventually the ability to run on a distributed network in order to speed up processing of large images. When MR PRISM is eventually released to beta test, we hope it will become the 'front end' for many more complicated routines from all branches of Mars research.

## 5. CONCLUSION

The overall goal of the MR PRISM project is to make the CRISM analysis more enjoyable, and hopefully reduce the barrier between hyperspectral analysis and the planetary science community.

In this paper we have demonstrated (for the first time) the MR PRISM software package, designed to analyze CRISM data on MRO. The potential of this application for easing the analysis task for the planetary science community is hopefully apparent. We would encourage any readers who are interested in participating in the beta testing stage of this software to contact the authors for further details. Beta testers who are positive, computer savvy and have a desire to make a difference for the planetary science community are sought to help make the project a success.

## ACKNOWLEDGEMENTS

AJB would like to acknowledge his postdoctoral adviser, Dr. Janice Bishop, for her assistance in obtaining the OMEGA data and for help with conceptualizing the design of MR PRISM. We would like to thank Dr. François Poulet, Dr. Jean-Pierre Bibring and the OMEGA team for providing the OMEGA data presented in this paper.

MR PRISM is only in the prototyping stage and already it owes much to the Java open source community. Development has been inspired by open source Java projects such as ImageJ, Jade Display and GeoVirgil. It uses the Java Advanced Imaging library from Sun Microsystems. Code has been used from open source projects such as: JGoodies, JFreeChart, Gui-Commands, Colt, GeoApi, Groovy and OsterMillerUtils. It was developed on the fantastic open source IDE, Eclipse.